\documentstyle[aps,prl,epsfig]{revtex}

\begin{document}
\draft

\title{ On The Detection of Scalar Field Induced Spacetime Torsion}

\author{T. Dereli\footnote{ Leverhulme Visiting Professor, on leave from the
Middle East Technical University, Ankara, Turkey} ,
R. W. Tucker}

\address{ Department of Physics, Lancaster University,\\
Lancaster LA1 4YB, UK \\ {\small t.dereli@lancaster.ac.uk}\\{\small
r.tucker@lancaster.ac.uk}}
\date{ 5 November 2000}

\maketitle

\begin{abstract}
\noindent We argue that the geodesic hypothesis based on autoparallels
of the Levi-Civita connection may need refinement in the
scalar-tensor theories of gravity. Based on a reformulation of
the Brans-Dicke  theory in terms of a connection with torsion determined
dynamically in terms of the gradient of the Brans-Dicke scalar
field, we compute the perihelion shift in the orbit of Mercury on
the alternative hypothesis that its worldine is an autoparallel of
a connection with torsion. If the Brans-Dicke scalar field couples
significantly to matter and test particles move on such worldlines, the
current time keeping methods based on the conventional geodesic
hypothesis may need refinement.
\end{abstract}


\def\pmb#1{\hbox{\bf #1}}
\def\bfr{{\pmb{r}}}
\def\bbfR{{\pmb{R}}}

\def\dotbfr{\dot{\bfr}}
\def\dotbbfR{\dot{\bfR}}
\def\dotbfR{\dot{\bfR}(\sigma,\eta)}

\def\^{\wedge}
\def\pd{\partial} 

\def\Lie{\hbox{\it \char'44}\!}

\def\bfT{{\bf {T}}}
\def\bfg{{\bf {g}}}

\def\bfd{\pmb{d}(\sigma,\eta)}
\def\bbfd{\pmb{d}}
\def\del{{ \nabla}}   

\def\LCdel{\hat \nabla}

\def\cs{ \left \{ {}^{\mu}_{\nu \lambda} \right \} }


Despite  continuing efforts to seek a fundamental description of
all the basic interactions in Nature, the role of gravitation in
this endeavor remains elusive. In part this may be due to our
inability to explore the effects of unified descriptions on scales
that are accessible to experiment. However, there are compelling
suggestions from astrophysical observations that Einstein's
original description of gravity may require the inclusion of
hitherto undetected  fields of either gravitational or matter
origin. Low energy effective string theories are replete with
unobserved scalar fields and most unified theories of the strong
and electroweak interactions predict fields with astrophysical
implications. In 1961 Brans and Dicke \cite{BD},\cite{brans}
suggested a modification of Einsteinian gravitation by introducing
an additional scalar field with particular gravitational couplings
to matter via the spacetime metric. This is arguably the simplest
modification and in this note we suggest that efforts to detect
such scalar gravitational interactions may have overlooked a
possibility that has experimentally detectable implications.

Relativistic gravitation benefits from a formulation in terms of
invariant structures on a spacetime manifold. Einstein's theory
takes as a fundamental field the spacetime metric tensor and
identifies gravitation with the spacetime curvature associated
with this metric. The effect of this curvature on matter fields is
recognised as due to the {\sl gravitational force} and it is commonly
assumed that massive point test particles have spacetime histories
that coincide with time-like geodesics associated with the
spacetime metric. In terms of the torsion-free connection
compatible with  this metric, such histories have self-parallel
tangent vectors and may be termed Levi-Civita autoparallels.
Assuming that our Sun generates an exterior Schwarzshild metric in
the vicinity of the planet Mercury, one may use this hypothesis to
calculate the perihelion shift per revolution of its orbit and
compare directly with observation. Despite competing perturbations,
this  prediction is regarded as one of the classical tests of any
theory of gravitation. Einstein, Infeld  and Hoffmann \cite{einstein}
 made valiant attempts
to prove this {\it geodesic hypothesis} for particles from a field
theory approach but their conclusions were not entirely
convincing. However, due to its naturalness, the geodesic hypothesis
for spinless test particles has almost become elevated to one of
the natural laws of physics \cite{weinberg}  and in
(pseudo-)~Riemannian spacetimes
it arises convincingly as the lowest approximation to a multipole
expansion of matter distributions in tidal interaction with
gravity \cite{papapetrou}, \cite{dixon}.
For massive test particles with spin and for all matter in
spacetimes with torsion, the foundation of this hypothesis is
less clear cut \cite{ponomariev}, \cite{trautman}.
Contrary to popular belief one does not require
spinor fields to generate spacetime torsion. If one takes the view
that the general setting for any description of gravitation is a
manifold equipped with a metric tensor field $\bfg$ and a
connection $\del$, then the question of the existence of torsion
depends on the connection. Connection forms (gauge potentials)
have proved the cornerstone in the unification of the strong and
electroweak interactions and it is natural to treat them as
independent dynamical entities in gravitation also. The dynamical
status of spacetime torsion is then dependent on the action chosen
for the theory with independent metric and connection fields.

For any vector fields $X$ and $Y$ on spacetime the torsion tensor
$\bfT$ is defined by: $$\bfT(X,Y)=\del_X\,Y -\del_Y\,X-[X,Y].$$ In
spacetimes with a (pseudo-)Riemannian  geometry one assumes
$\bfT=0$ and $\del \bfg=0 $ and the connection is given in terms
of the Christoffel symbols $\cs$ obtained by differentiating
components of the metric tensor. To distinguish this Levi-Civita
connection from a general one we denote it $\LCdel$.

In the scalar-tensor theory of  gravitation formulated by Brans
and Dicke \cite{BD}, the motion of a test particle was originally
assumed to be a Levi-Civita geodesic associated with the metric
derived from the Brans-Dicke field equations (even though the
scalar field could vary in spacetime). Later Dirac in \cite{Dirac}
showed that in a Weyl invariant generalisation it was more natural
to generate the motion of a test particle from a Weyl invariant
action principle and that such a motion in general differed from a
Brans-Dicke Levi-Civita geodesic. Although Dirac was concerned
with the identification of electromagnetism with aspects of Weyl
geometry even for neutral test particles it turns out that test
particles would follow auto-parallels of a connection with
torsion. In Ref. \cite{DT} we have shown that the theory of Brans
and Dicke \cite{BD} can be reformulated as a field theory on a
spacetime with dynamic torsion determined by the gradient of the
Brans-Dicke scalar field $\Phi$. Of course no new physics {\it of
the fields} can arise from such a reformulation, although it does
clarify certain issues relating to the conformal structure of the
theory and its couplings to matter with intrinsic spin. However,
the behaviour of spinless {\it particles} in such a geometry with
torsion depends on the choice made from two possible alternatives.
One may assert that their histories are {\sl either} geodesics
associated with autoparallels of the Levi-Civita connection {\sl
or}  the autoparallels of the non-Riemannian connection with
torsion. Since one may find a spherically symmetric, static
solution to the Brans-Dicke theory (in either formulation), it is
possible to compare these alternatives for the history of Mercury
about the Sun by regarding it as a spinless test particle as in
General Relativity. In this letter we report on the results of
this computation.

\bigskip

In terms of local coordinates $\{t,\rho,\theta, \phi\}$ the
spherically symmetric static solution \cite{BD} in the vicinity of
Mercury is taken in the form:

\def\LM{1-\frac{\Lambda}{\rho}}
\def\LP{1+\frac{\Lambda}{\rho}}

\def\gtt{\left (\frac{\LM}{\LP}\right )^{2p}}

\def\gss{ \left (\LP\right )^4 \left (\frac{\LM}{\LP}\right )^{2-2p-2q}   }

$$\bfg=-\gtt c^2 d\,t \otimes d\,t + \gss (d\,\rho \otimes d\,\rho
+\rho^2\, d\,\theta\otimes d\,\theta\ + \rho^2\,\sin^2\theta
d\,\phi\otimes d\,\phi)$$ $$\Phi=\Phi_0\left
(\frac{\LM}{\LP}\right )^q$$ where, in terms of the conventional
Brans-Dicke coupling parameter $\omega$, \, $p=\left (
\frac{2\omega+4}{2\omega+3}  \right )^{1/2}$, \, $q=p\,(\gamma
-1)$ with $\gamma=\frac{\omega+1}{\omega+2}$. \,  The constants
$\Lambda$ and $\Phi_0$ are fixed by ensuring that in the weak
field limit \cite{BD} one can identify the Newtonian coupling
constant $G$ and the source of matter as the solar mass $M$:
$$\Lambda=\frac{GM}{2c^2}\left (\frac{2\omega+3}{2\omega+4} \right
)^{1/2}$$ $$\Phi_0=\frac{1}{G}\frac{2\omega+4}{2\omega+3}.$$ In
order to relate this solution to a frame more appropriate to
terrestrial observation, we effect the transformation
$$r=\rho\left (\LP \right )^{1+p+q} \left (\LM \right )^{1-p-q}$$
and express the metric as
\begin{eqnarray} \bfg=-A(r) c^2 d\,t \otimes d\,t + B(r)^{-1} d\,r
\otimes d\,r + r^2\,( d\,\theta\otimes d\,\theta\ +
\,\sin^2\theta d\,\phi\otimes d\,\phi)\label{approxg}
\end{eqnarray}
where $$A(r)=1-\frac{2GM}{c^2r}+a_2\frac{(2GM)^2}{c^4
r^2}+\ldots$$ $$B(r)=1+b_1\frac{2GM}{c^2 r}+\ldots$$
$$\Phi(r)=\Phi_0 \left (1+c_1\frac{2GM}{c^2r}+\ldots \right )$$ with
$a_2=c_1=\frac{1-\gamma}{2}$, $b_1=-\gamma$.
\def\Cdot{{\dot C}}
In the geometry with torsion, the torsion 2-forms defined by
$T^a(X,Y)=\frac{1}{2}e^a(\bfT(X,Y))$ with \, $\bfg= - e^0 \otimes
e^0 + \sum^{3}_{j=1}e^j\otimes e^j$ \, are given by \cite{DT}:
$$T^a=e^a\^\frac{d\,\Phi}{2\Phi}.$$ The equation for a timelike
autoparallel is
 $$\del_\Cdot\,\Cdot=0$$
where the 4-velocity $\Cdot$ is normalised with
\begin{eqnarray}
\bfg(\Cdot,\Cdot)=-c^2\label{norm}.
\end{eqnarray}
By expressing $\del$ in terms of the Levi-Civita connection
$\LCdel$ with $\tilde{V}=g(V,-)$ for any vector $V$ one may write
this as
$$\widetilde{\LCdel_\Cdot\Cdot}=-\frac{1}{2\Phi}i_{\Cdot}(d\,\Phi\^\tilde{\Cdot})
$$ (the operator $i_\Cdot$ denotes contraction of the 2-form with
the vector $\Cdot$ )
 and interpret the right hand side as a torsion acceleration
field analogous to the Lorentz force on electrically charged
particles. Note however that the torsion force  produces the same
acceleration on all  massive  test particles. If $\Cdot$ is
parameterised in terms of proper time $\tau$ in any coordinates
$x^\mu(\tau)$, the above is:
\begin{eqnarray}
\frac{d}{d\,\tau}\left (\Phi^{1/2}\frac{d\,x^\mu}{d\,\tau} \right )+\Phi^{1/2}
\cs
\frac{d\,x^\nu}{d\,\tau}\frac{d\,x^\lambda}{d\,\tau}=-g^{\mu\nu}\frac{\partial_\nu\Phi}{2\Phi^{1/2}}.
\label{auto}
\end{eqnarray}
(This equation coincides with equation (8.8) in \cite{Dirac} for a
neutral test particle in Dirac's reformulation of Weyl's theory
mentioned above. The apparent sign difference is due to
signature conventions.) 
) To solve this equation one notes that for any
Killing vector $K$, ($\Lie_K \bfg=0$) with $K\,\Phi=0$,
expressions proportional to $\Phi^{1/2}\bfg(K,\Cdot)$ are constant
along the test particle worldline. For the above solution
(\ref{approxg}), the Killing vectors  $K_t=\partial_t$ and
$K_\phi=\partial_\phi$ generate two proper time  constants of the
motion $E$ and $L$ corresponding to energy and angular momentum
about any direction in space, respectively. Choosing solutions
with $\theta=\pi/2$, we set
\begin{eqnarray}
E=-mc\left (\frac{\Phi}{\Phi_0}\right )^{1/2}A(r)\dot
t\label{energy}
\end{eqnarray}
\begin{eqnarray}
L=m \left (\frac{\Phi}{\Phi_0}\right
)^{1/2}r^2\dot \phi.\label{amom}
\end{eqnarray}
 To facilitate
\def\EE{{\cal E}}
comparison with non-relativistic Kepler orbits we write
$E^2=(mc^2+\EE)^2$ in terms of the (constant) rest mass $m$ of
Mercury and identify $\EE$ with its (negative) gravitational
binding energy. Similarly let $L^2=m^2 h^2$ with $h=\sqrt{GMr_0}$.
Then the standard Newtonian Kepler orbit arises from $\Phi=\Phi_0$
above and is given in the non-relativistic limit by  the geodesic
hypothesis as $$ \frac{1}{r}=\frac{1}{r_0}(1+\epsilon \cos\phi)$$
where the eccentricity for a bound orbit is
$\epsilon=\sqrt{1+\frac{2\EE r_0}{GMm}}$.

From (\ref{norm}) one may eliminate $\dot t$ and write the
equation for the planar orbit of Mercury according to
(\ref{auto}). Taking into account that the speed of Mercury is
non-relativistic and that its Newtonian orbit is much larger than
the Schwarzschild radius $r_s=2GM/c^2$ of the Sun, one finds in
terms of $U(\phi)=1/r(\phi)$ the orbit equation

$$ \left (\frac{d\,U}{d\,\phi} \right )^2 \simeq   \frac{2\EE}{GMmr_0} +
\left (\frac{2}{r_0}-\frac{1}{r_0(\omega+2)}\right )U -U^2 $$
\begin{eqnarray}
+\frac{(1-a_2+b_1-b_1c_1)}{L^2}\frac{(2GMm)^2U^2}{c^4}- \frac{2 G
M b_1U^3}{c^2}. \label{diffeqn}
\end{eqnarray}
We note that the last two terms on the right hand side of this
\def\rh0{\hat r_0}
\def\eh0{\hat\epsilon}
equation yield the leading corrections to the non-relativistic
Newtonian orbit. The latter however also receives contributions
from the Brans-Dicke interaction. The Newtonian orbit is the
Keplerian ellipse:
 $$\hat U(\phi)=\frac{1}{\rh0}(1+\eh0\cos\phi)$$
with $$\rh0=r_0\frac{2\omega+4}{2\omega+3}$$ $$\eh0^2=1+\frac{8\EE
r_0}{GMm}(\frac{\omega+2}{2\omega+3})^2.$$
 Equation (\ref{diffeqn})  can be integrated exactly
in terms of Jacobi elliptic functions.

A straightforward analysis of the periodicity of such solutions
enables one to compute  the perihelion shift per revolution
$\Delta$ of the orbit:
\begin{eqnarray}
\Delta=\frac{3\omega+5}{3\omega+6}\,\,\delta_\omega
\label{ans1}
\end{eqnarray}
where $\delta_\omega=3\lambda_\omega\pi$ and
$\lambda_\omega={r_s}/{\rh0}$. Using the Kepler period $\hat
T=2\pi({\rh0^3}/{(1-\eh0^2)GM})^{1/2}$, the shift $\delta_\omega$
may be expressed in terms of $\hat T$ and $\eh0$. This may be
compared with the result based on the assumption that Mercury's
orbit follows from a geodesic of the torsion-free Levi-Civita
connection:
 $$\LCdel_\Cdot\Cdot=0.$$ In this case one
finds
\begin{eqnarray}
\Delta=\frac{3\omega+4}{3\omega+6}\,\,\delta\label{ans2}
\end{eqnarray}
where $\delta=3\lambda\pi$ with $\lambda={r_s}/{{r_0}^\prime}$
 is the  perihelion shift
per revolution of the orbit based on the Schwarzschild solution
for the metric in General Relativity \cite{will}.

\vskip 0.4cm

 \centerline{\bf Constraints on $\omega$ imposed by
Observation of the Planet Mercury }

\begin{figure}
\begin{center}
\epsfig{file=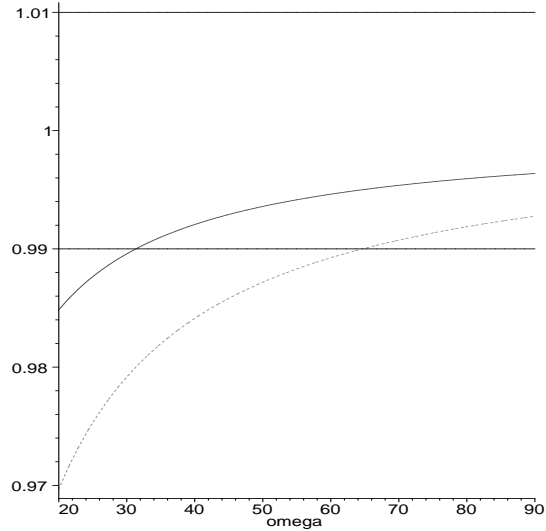,width=70mm,height=70mm}

 \vskip 0.4cm
 \caption{Behaviour of $\Delta/\delta$ as a function
of $\omega$. The full curve  corresponds to a precession rate of
Mercury's orbit under both metric and torsional acceleration. The
dotted line corresponds to the original prediction of the
Brans-Dicke theory. The experimental observations are consistent
with $\Delta/\delta$ lying between the full horizontal lines
centered on $\Delta/\delta=1$ }
\end{center}
\end{figure}

It should be noted that $r_0^\prime$ is the parameter of a
Newtonian ellipse  with   Kepler period $T=2\pi(
{{r_0^\prime}^3}/{(1-\epsilon^2)GM})^{1/2}$ and eccentricity
$\epsilon$. However in terms of the energy and angular momentum of
the particle in orbit the relations (\ref{energy}) and
(\ref{amom}) are replaced by similar ones but with $\Phi=\Phi_0$.

In order to compare the prediction (\ref{ans1}) with that of
Brans-Dicke (\ref{ans2}), we recognise that the orbit parameters
must be common (i.e. constants of motion choosen so that
$\rh0=r_0^\prime$). Thus with a common $\delta$ we may plot
$\Delta/\delta$ as a function of $\omega$ in a domain where both
enter the error corridor of this value determined by observation.
According to \cite{will} the above corridor can probably be
reduced by half although we shall be conservative in our error
estimates.  The differences indicated in the figure suggests that
in assessing the significance of the scalar field in the
Brans-Dicke description of gravity, one should take seriously the
possibility that Mercury's orbit might be described by an
autoparallel of the natural connection with torsion used in the
alternative formulation of the theory.
Given the recent
improvements in satellite technology and space location techniques,
a more reliable method to detect the effects of torsion induced
motion would be to measure the precession of satellites in highly
eccentric orbits about spherical asteroids. Indeed, if the
Brans-Dicke scalar field couples significantly to matter  and test
particles move on autoparallels of connections other than the one
in General Relativity, current time keeping methods based on the
conventional geodesic hypothesis may need refinement.


\bigskip


\noindent {\bf
Acknowledgement}

RWT is  grateful to Middle East Technical University for its
hospitality and to the Scientific and Technical Research Council
of Turkey (TUBITAK), BAE-Systems (Warton) and  PPARC for financial
support. Both authors are grateful for support from the Leverhulme
Trust.



\vfil\eject

\end{document}